\documentclass[twocolumn]{aastex62}
\usepackage{amssymb}
\usepackage{amsmath}
\usepackage{amssymb}
\usepackage{empheq}
\usepackage{mathrsfs}
\usepackage{wasysym} 

\usepackage{etoolbox}
\usepackage{lmodern}

\newcommand{\be}{\begin{equation}}
\newcommand{\ee}{\end{equation}}

\begin{document} 

\title{\bf Generation of Low-Inclination, Neptune-Crossing TNOs by Planet Nine}

\author[0000-0002-7094-7908]{Konstantin Batygin}
\affiliation{Division of Geological and Planetary Sciences, California Institute of Technology, Pasadena, CA 91125}

\author{Alessandro Morbidelli}
\affiliation{Laboratoire Lagrange, Universit\'e C\^ote d'Azur, Observatoire de la C\^ote d'Azur, CNRS, CS 34229, F-06304 Nice, France}

\author[0000-0002-8255-0545]{Michael E. Brown}
\affiliation{Division of Geological and Planetary Sciences, California Institute of Technology, Pasadena, CA 91125}

\author[00000-0002-4547-4301]{David Nesvorn\'y}
\affiliation{Department of Space Studies, Southwest Research Institute, 1050 Walnut St., Suite 300, Boulder, CO 80302}

\begin{abstract}
The solar system's distant reaches exhibit a wealth of anomalous dynamical structure, hinting at the presence of a yet-undetected, massive trans-Neptunian body — Planet 9. Previous analyses have shown how orbital evolution induced by this object can explain the origins of a broad assortment of exotic orbits, ranging from those characterized by high perihelia to those with extreme inclinations. In this work, we shift the focus toward a more conventional class of TNOs, and consider the observed census of long-period, nearly planar, Neptune-crossing objects as a hitherto-unexplored probe of the Planet 9 hypothesis. To this end, we carry out comprehensive $N-$body simulations that self-consistently model gravitational perturbations from all giant planets, the Galactic tide, as well as passing stars, stemming from initial conditions that account for the primordial giant planet migration and sun’s early evolution within a star cluster. Accounting for observational biases, our results reveal that the orbital architecture of this group of objects aligns closely with the predictions of the P9-inclusive model. In stark contrast, the P9-free scenario is statistically rejected at a $\sim5\,\sigma$ confidence-level. Accordingly, this work introduces a new line of evidence supporting the existence of Planet 9 and further delineates a series of observational predictions poised for near-term resolution.
\end{abstract}


\section{Introduction}
\label{sec:intro} 

The discovery and characterization of the trans-Neptunian population of small bodies have played a pivotal role in the reimagining of the narrative of our solar system's long-term evolution. Beyond a qualitative shift towards an instability-driven scenario (commonly referred to as the Nice model; \citealt{2005Natur.435..459T, 2005Natur.435..462M, 2005Natur.435..466G}), detailed modeling of the Kuiper belt's formation has brought the migratory histories of the giants planets into a remarkable degree of focus \citep{NesvReview}. As advancements in observational surveys have sharpened our understanding of the outer solar system's orbital architecture, however, a series of anomalous patterns that cannot readily be attributed to early dynamical sculpting, have been unveiled.

These anomalies include the apparent clustering of apsidal lines of long-period trans-Neptunian object (TNO) orbits, the alignment of their orbital planes, the existence of objects with perihelia extending far beyond Neptune's gravitational influence, the highly extended distribution of TNO inclinations, and the surprising prevalence of retrograde Centaurs. Collectively, these irregularities hint at the existence of a yet-undiscovered massive planet, tentatively named Planet Nine (P9), whose gravitational influence sculpts the outer reaches of trans-Neptunian space \citep{B19}. While these patterns were largely identified in a series of papers dating back 8 years or more \citep{2004ApJ...617..645B, 2009ApJ...697L..91G, 2014Natur.507..471T, 2015Icar..258...37G, BB2016a}, numerous studies carried out over the last decade have explored how the dynamical influence of P9 could shape the solar system's observed characteristics \citep{BB2016a, BB2016b, BroBat16, Beust2016, BM17, 2017AJ....153...91M, 2017AJ....154...61B, 2017CeMDA.129..329S, 2018AJ....155..249H, 2018AJ....156..263L, 2018AJ....156...81B, B19, Kaib19, 2020PASP..132l4401K, 2020AJ....159..285C, 2021AJ....162...39O, 2021AJ....162...27C, BB21, 2021AJ....162..219B}.

\begin{figure*}
\centering
\includegraphics[width=\textwidth]{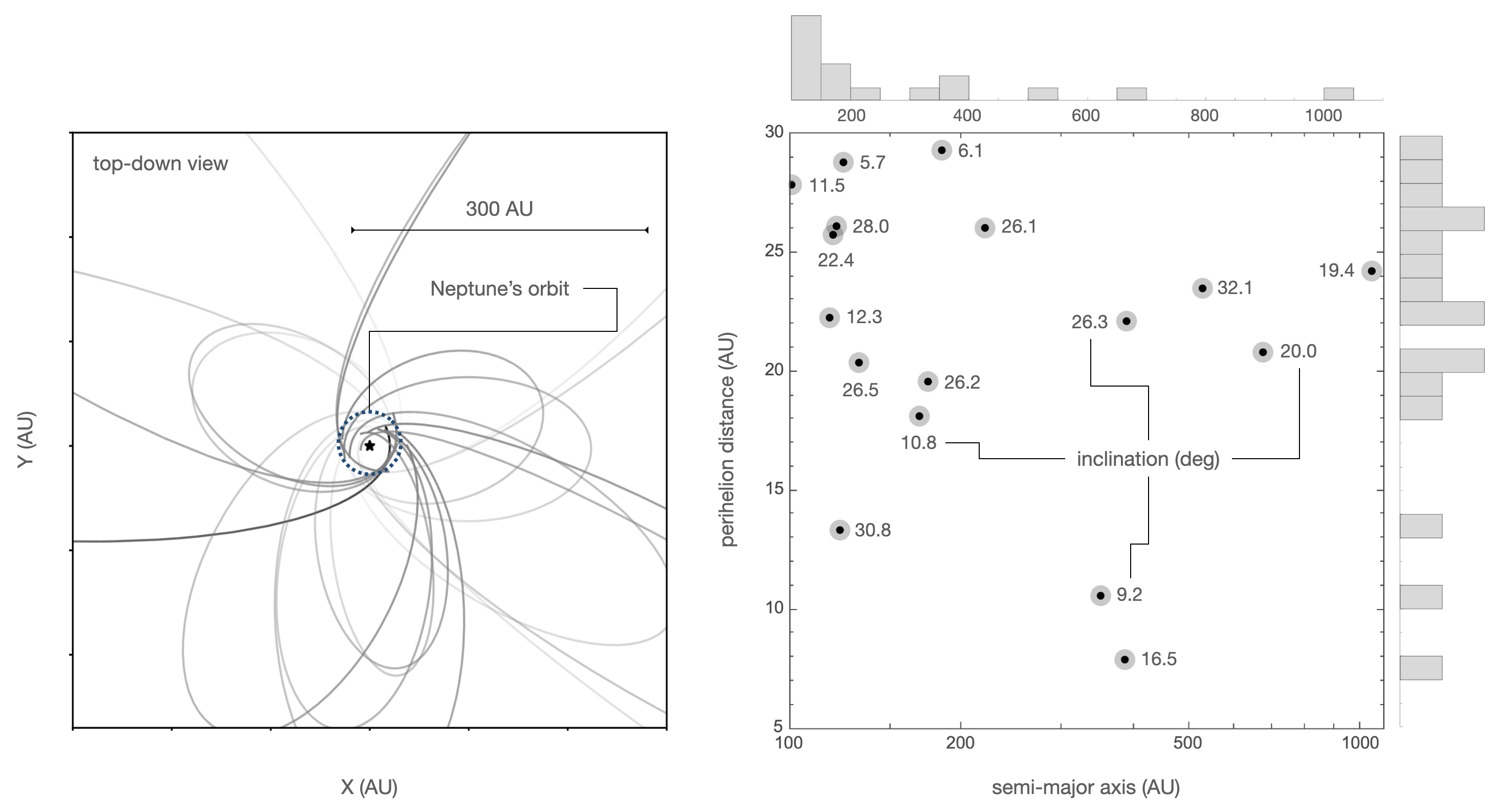}
\caption{Census of well-characterized TNOs with $a>100\,$AU, $i<40\deg$, and $q<30\,$AU. Among the 29 objects within the Minor Planet Center database that meet these orbital criteria, our analysis is restricted to 17 objects whose orbits have been quantified through multi-opposition observations.} The left panel shows a top-down view of the orbits. The right panel depicts a plot of perihelion distance against semi-major axis; the numbers adjacent to the points indicate each object's orbital inclination in degrees. Notably, the spread of perihelion distances forms a relatively flat distribution between $\sim16\,$AU and Neptune's orbit.
\label{fig:data}
\end{figure*}

Generally speaking, the observed irregularities among trans-Neptunian objects can be categorized into those relating to dynamically detached objects (such as the clustering of longitudes of perihelion and alignment of orbital planes) and those associated with chaotic, Neptune-crossing orbits, particularly evident in the highly-inclined population. Under the Planet Nine hypothesis, however, the boundary between these categories is somewhat blurred, since dynamical evolution driven by P9 can cause long-period TNOs to oscillate between detached and Neptune-crossing states over secular ($\sim$Gyr) timescales (e.g., \citealt{B19}). This simple fact necessitates an important consequence: if Planet Nine exists, it should continuously produce nearly planar ($i < 40\deg$), long-period ($a > 100\,$AU) objects with perihelia smaller than $q<30\,$AU. Remarkably, more than a dozen multi-opposition objects fitting this description have been identified (Figure \ref{fig:data}), yet their significance within the context of the Planet Nine hypothesis remains unexplored.

A principal goal of this study is to analyze the dynamical origins of these objects to assess their potential in serving as a new probe for Planet 9. To this end, we carry out two sets of comprehensive numerical simulations: one considering the gravitational influence of Planet 9 -- where the generation of nearly-planar, long-period, low-$q$ orbits is facilitated by P9's gravity -- and the other excluding it, where the evolution of distant TNOs is driven primarily by Neptune-scattering and the Galactic tide \citep{2014Icar..231..110F}. Together, these numerical experiments demonstrate that, while Neptune constitutes a veritable barrier for scattered disk objects, P9-driven evolution allows for perturbed orbits to readily cross this threshold, creating a distinct signature. Moreover, upon accounting for observational bias using a novel approach, our calculations show that the distribution of the observed orbits strongly supports the presence of the unseen planet. The remainder of this paper is organized as follows. Section 2 outlines our numerical simulation methods. In Section 3, we present our findings, discuss the treatment of observational biases, and compare our results with both observational data and a state-of-the-art P9-free model of the solar system's architecture. Our conclusions and further observational predictions stemming from our calculations are discussed in Section 4.

\section{Numerical Simulations}
\label{sec:numsim} 

Since its inception, numerical modeling of the Planet Nine hypothesis has varied significantly in complexity, ranging from simplified, orbit-averaged descriptions of the dynamics to more detailed simulations that include Planet Nine and Neptune or all giant planets as active perturbers. In this work, we adopt the latter approach and further incorporate extrinsic effects from our previous study \citep{BB21}, which self-consistently included the effects of passing stars and the galactic tide. We describe our numerical setup in detail below.

\paragraph{Simulated Interactions} Given our focus on resolving the evolution of objects with perihelia smaller than 30 AU, any form of orbital averaging is unsuitable for our purposes. Therefore, our simulations include Jupiter, Saturn, Uranus, and Neptune, initialized on their present-day orbits, as active perturbers. For Planet Nine, we adopt a mass of 5 Earth masses, placing it in an orbit with a semi-major axis of 500 AU, an eccentricity of 0.25, and an inclination of 20 degrees. While the precise orbit of Planet Nine remains unknown \citep{2021AJ....162..219B}, this configuration aligns with previous studies and satisfactorily accounts for the orbital anomalies outlined in the Introduction (namely, clustering of the longitudes of perihelion, grouping of the orbital poles, etc. -- see \citealt{B19} and the references therein). While a full exploration of Planet Nine's parameter space is beyond the scope of this study, it is likely that any combination of P9 parameters that results in significant orbital clustering among perihelion-detached orbits will also produce commensurate effects for Neptune-crossing orbits, since both effects are driven by secular eccentricity modulation.

In addition to planetary perturbations, our simulations incorporate Galactic effects. Galactic tidal accelerations are included following a standard approach (see \citet{2017ApJ...845...27N} and references therein), and passing stars are introduced following the procedure described in \citet{1986Icar...65...13H}. While these effects are generally weak for orbits with $a<1000\,$AU, they can play an important role for objects that diffuse outwards to much larger semi-major axes, before being scattered back to shorter orbital periods by Neptune. We do not consider the possibility of the Sun's radial migration through the galaxy \citep{2011Icar..215..491K} and neglect the self-gravity of the Oort Cloud \citep{BN24} for definitiveness.

\paragraph{Initial Conditions} The influence of initial conditions in P9 simulations was first demonstrated by \citet{Kain18}, who showed that a broadened initial perihelion distribution is generally preferable to a narrow one. Importantly, this broadening is expected, given that the solar system almost certainly originated within a star cluster (see \citealt{Adams2010, 2023A&A...670A.105A}), and the orbital distribution of TNOs would have been affected by cluster dynamics within the first $\sim100\,$Myr of the solar system's evolution. Following this reasoning, in \citet{BB21}, we generated initial conditions by simulating the formation of the inner Oort Cloud, while accounting for constraints emanating from the orbital structure of the cold classical belt \citep{B20}. In a recent study, \citet{Nesv23} modeled the formation of the primordial trans-Neptunian population using equivalent cluster parameters, while also accounting for a comprehensive description of the early orbital migration of the giant planets \citep{NesvReview}.

Here, we adopt the $t=300\,$Myr timestamp from the \texttt{cluster\_2} simulation of \citet{Nesv23} as our starting point, treating all TNOs as massless test particles. This epoch is early enough that the intrinsic dynamical evolution in the outer solar system is still in its infancy, but late enough that the solar system's birth cluster would have already dispersed, and the  migration of the giant planets largely concluded. Within this synthetic dataset of $\sim10^5$ objects, our particle selection encompasses bodies with perihelia greater than 30 AU (that is, we remove all Neptune-crossing objects from the initial conditions) and semi-major axes between 100 and 5000 AU, with a total count of approximately 2000 particles. Objects interior to $\sim100\,$AU are not strongly influenced by Planet 9, while bodies outside of $5000\,$AU are both relatively sparse, and achieve large enough heliocentric distances to be dominated by Galactic effects.

Importantly, this choice of initial conditions is inherently linked with the assumed orbit of Planet 9. Arguably the most plausible origin scenario for Planet 9 involves formation within the protosolar nebula, followed by outward scattering by Jupiter and Saturn. This process necessitates strong stellar perturbations to effectively detach P9's orbit from those of the giant planets (essentially rendering P9 itself an Inner Oort Cloud object; \citealt{B19,2023DPS....5550009I}). While the \texttt{cluster\_2} simulation of \citet{Nesv23} readily generates detached orbits akin to the one we assumed for P9, the \texttt{cluster\_1} simulation -- which is characterized by weaker stellar perturbations -- does not\footnote{The most analogous orbit to our assumed parameters for P9 within the \texttt{cluster\_2} simulation has $a=541\,$AU, $e=0.32$, and $i=20\deg$. In contrast, the closest orbit generated in the \texttt{cluster\_1} simulation has a $a=708\,$AU, $e=0.45$, and $i=25\deg$.}. This discrepancy underscores the necessity of a relatively densely populated stellar environment for self-consistently achieving the orbital parameters we assume for P9.

\paragraph{Integration Method} To carry out the integrations, we used the conservative variant of the Bulirsch-Stoer algorithm, as implemented in the \texttt{mercury6} gravitational dynamics software package \citep{1999MNRAS.304..793C}. The initial timestep was set to 100 days, but was altered adaptively, satisfying an accuracy parameter of $\epsilon = 10^{-11}$ \citep{1992nrfa.book.....P}. The simulation's inner and outer absorbing boundaries were defined at 1 and 100,000 AU respectively, with passing stars introduced at the exterior boundary. The integration was carried out over a timespan of $4\,$Gyr.

\begin{figure}
\centering
\includegraphics[width=\columnwidth]{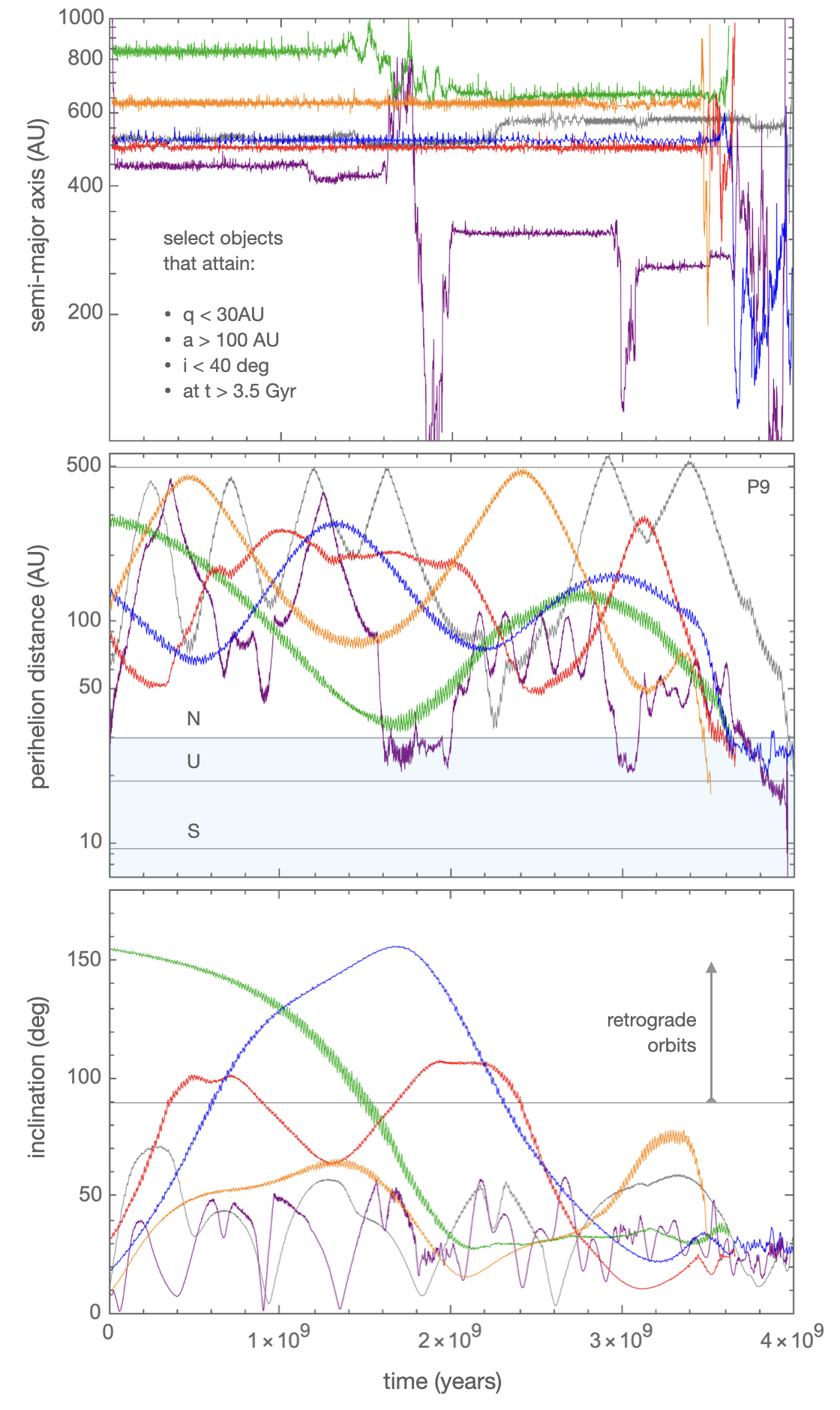}
\caption{Evolution of selected particles within our calculations that attain nearly planar ($i<40\deg$) Neptune-crossing orbits, within the final $500\,$Myr of the integration. The top, middle and bottom panels depict the time series of semi-major axis, perihelion distance, and inclination, respectively. Some particles experience large perihelion oscillations while remaining on prograde orbits for the duration of the simulation. Others exhibit coupled eccentricity-inclination dynamics the drive orbital flips. Although the orbital evolution is always stochastic, the rate of chaotic diffusion greatly increases when particles attain Neptune-crossing trajectories.}
\label{fig:timeseries}
\end{figure}

\begin{figure*}
\centering
\includegraphics[width=\textwidth]{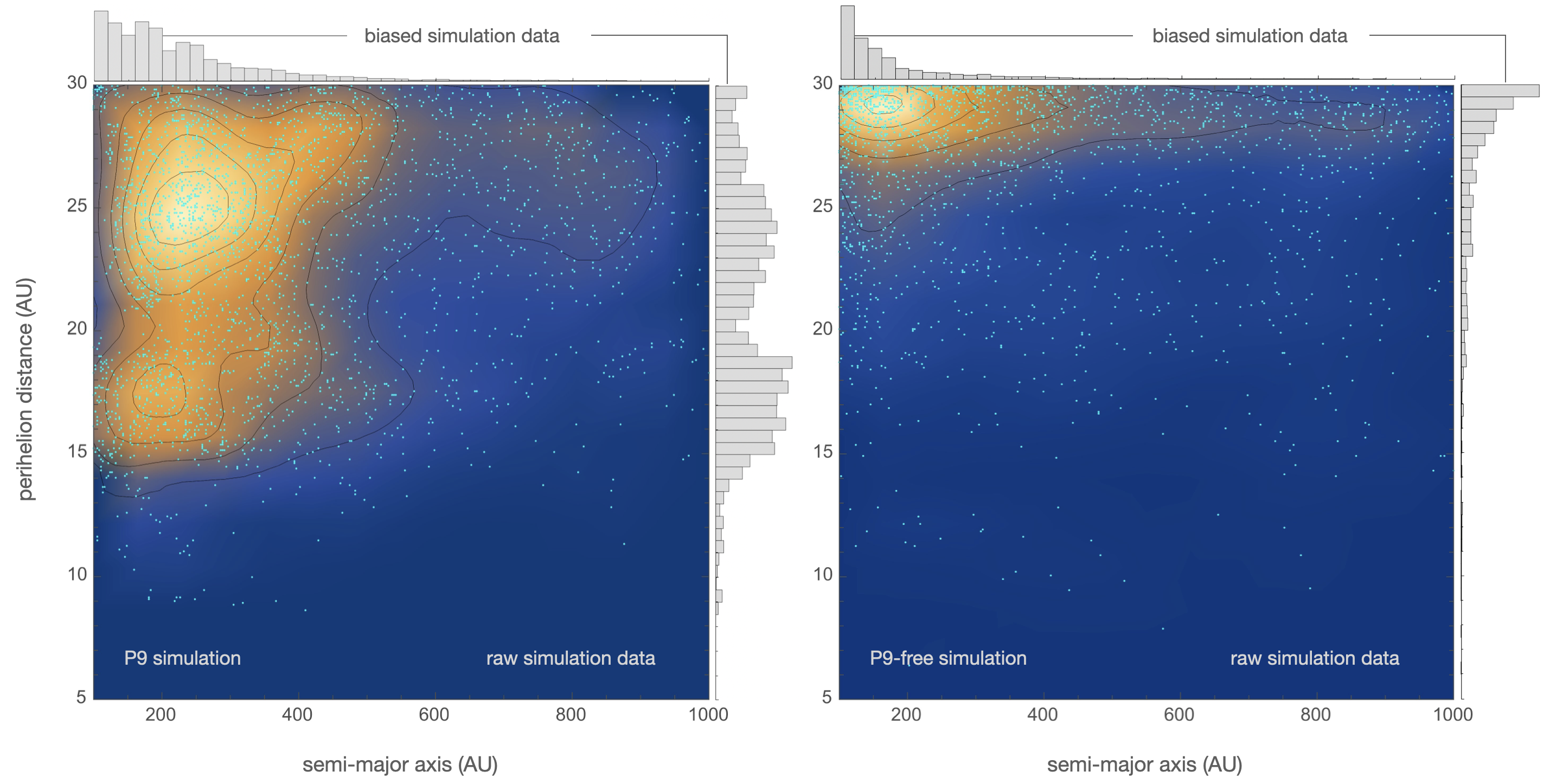}
\caption{A comparison of the orbital distributions from P9-inclusive (left) and P9-free (right) $N-$body simulations. Both panels depict the perihelion distance against the semi-major axis of orbital footprints of simulated TNOs with $i<40\deg$. The overlaying contour lines represent density distributions, with brighter colors indicating higher concentrations of objects. While the panels themselves show raw simulation data, the histograms along the axes show a biased frequency distribution for the perihelion distances (vertical) and semi-major axes (horizontal), assuming a limiting magnitude of $V_{\rm{lim}}=24$.} 
\label{fig:orbdist}
\end{figure*}

\smallskip

\section{Results}

Perihelion oscillations of detached TNOs under P9's influence are primarily driven by two secular effects. The first is direct Runge-Lenz vector coupling, akin to that captured by Lagrange-Laplace secular theory but occurring at high eccentricity \citep{Beust2016}. The second is a mixed inclination-eccentricity interaction (not to be confused with the von Zeipel-Lidov-Kozai effect, which is a secular resonance in the argument of the pericenter) that is driven by an octuple-order harmonic, $2\,\Omega-\varpi-\varpi_9$ \citep{BM17}. Our simulations reveal that both effects contribute to generating Neptune-crossing orbits. Figure \ref{fig:timeseries} displays the $a, e, i$ time-series of selected particles that achieve long-period, nearly planar orbits with $q<30\,$AU, within the final 500 Myr of the integration. Once a Neptune-crossing state is attained, the evolution becomes highly chaotic, marked by rapid random walk of the semi-major axis. It is notable, however, that this stochasticity does not invariably lead to ejection; trajectories can return to the scattered disk, or even undergo subsequent perihelion-detachment, as illustrated by the orbit shown in purple on Figure \ref{fig:timeseries}.

Collectively, these examples indicate that P9-facilitated dynamics can naturally produce objects similar to those depicted in Figure \ref{fig:data}. Still, the mere presence of such bodies in simulations is by no means sufficient as evidence for Planet 9. These objects could also, in principle, be generated by the combined action of Neptune-scattering and the Galactic tide, even in the absence of P9 \citep{1996CeMDA..64..209T,1999Icar..137...84W}. Thus, to more accurately assess the role of Planet 9, we focus on the perihelion \textit{distribution} of these low-inclination, Neptune-crossing orbits. As we will see below, the characteristics of this distribution provide a discerning diagnostic for P9-driven dynamics.

\subsection{Orbital Distributions}

To construct the inter-Neptunian perihelion distribution, we followed previous studies (e.g., \citealt{2018AJ....156..263L, B19, 2018AJ....155..249H, 2018AJ....156...81B, B19, 2021AJ....162..219B} and references therein) and examined the orbital footprints generated by particles satisfying the same orbital cuts as those adopted in Figure \ref{fig:data}, within the final Gyr of the integration. These footprints were recorded at 1 Myr intervals, far exceeding the typical Lyapunov time of the particles\footnote{Deep within the chaotic layer, the Lyapunov time of scattering objects approaches the orbital period \citep{2021ApJ...920..148B}.}. Due to chaotic mixing, any two footprints, even if sequentially produced by the same particle, are effectively uncorrelated. 

As a null hypothesis, we considered the P9-free \texttt{cluster\_2} simulation of \citet{Nesv23}. Despite having been conducted with a different integrator, this simulation includes all of the physical effects described in Section 2, with nearly identical implementation. Moreover, the \citet{Nesv23} model, previously validated against the distribution of high-$q$ TNOs, represents the current benchmark for the post-nebular evolution of the solar system. Given the larger particle count in this simulation compared to our P9 model, we sampled the final Gyr at $20\,$Myr intervals, yielding a similar number of footprints (between two and three thousand) satisfying $a>100\,$AU, $q<30\,$AU, $i<40\deg$, and a sufficiently large sample to construct smooth histograms for both scenarios. We further verified that the histogram shapes remained consistent over time, indicating that the flux of Neptune-crossing objects had attained steady-state by the final Gyr of the integration.

The left and right panels of Figure \ref{fig:orbdist} compare the raw (unbiased) $i<40\deg$ semi-major axis-perihelion distributions from simulations with and without P9, respectively. While both models yield semi-major axis distributions that diminish with increasing $a$ at long periods, the perihelion distributions are markedly different. The P9-free run shows a rapid decline in perihelion distribution with decreasing $q$, as Neptune's orbit forms a veritable dynamical barrier. In contrast, the simulation that includes P9 results in a relatively flat $q-$distribution outside $\sim16\,$AU, with a notable dip at $q\sim20,$AU (this feature can be attributed to strong gravitational interactions with Uranus, leading to a mild depletion in the perihelion distance distribution at this specific range).

Qualitatively, these differing $q-$distributions are tenable. In the P9-free scenario, objects with $a\lesssim1000\,$AU are too close to be significantly influenced by Galactic effects, leaving chaotic diffusion — which tends to preserve $q\sim a_{\text{N}}$ — as the primary driver of orbital evolution. On the other hand, P9-induced dynamics can continuously modulate the perihelion, even if the orbit dips well below Neptune's semi-major axis. Although the existing observational data indeed reveals a perihelion distribution that is relatively flat (Figure \ref{fig:data}), we cannot compare the data to the modeled distributions without accounting for observational bias.

\subsection{Correcting for Observational Bias}

Well-known observational biases exist against detecting orbits with high inclinations, as well as objects at large heliocentric distances. The former effect arises because observational surveys -- such as Pan-STARRS-1 and -2, which account for the discovery of a significant fraction of the observed objects -- are often performed at low ecliptic latitudes, where high inclination objects spend less time. Low-$i$ objects are thus over-represented in the catalogs. Fortunately, for the problem at hand, this bias is largely inconsequential because the numerical models do not show substantial difference in the perihelion distribution of $q<30\,$AU objects as a function of inclination, in the $i\sim0-40\deg$ range\footnote{This is not the case at significantly higher inclinations: above $i\gtrsim60\deg$, a population of large-$a$ Neptune-crossers also develops in the numerical models.}. Thus, by restricting our analysis to TNOs with inclinations lower than 40 degrees, we mitigate the principal source of inclination bias. Addressing the heliocentric distance bias, on the other hand, requires a more nuanced treatment, which is developed below.

\paragraph{Bias correction informed by discovery distance.} To account for the bias towards detecting objects with lower perihelia, we begin by examining the heliocentric distance at which each of the 17 known objects depicted in Figure \ref{fig:data}, were discovered. To leading order, at the moment of discovery, each object serves as an unbiased probe of the entire perihelion distribution, up to its discovery distance. This notion effectively nullifies biases related to discovery distance or object size, as the increasing brightness of an object with diminishing heliocentric distance becomes irrelevant. 

As a concrete example, consider a large sample of objects, all detected at $30\,$AU. To first approximation, the perihelia of this collection of bodies constitutes an unbiased probe of the $q-$distribution inside of $30\,$AU, regardless of the brightness of the object at the time of discovery\footnote{Strictly speaking, this statement assumes that the orbital distribution of TNOs is independent of their size.}, or the limiting magnitude of the survey. To advance beyond this estimate, one important correction must be made: objects with different $q$ and $a$ spend different fractions of their orbital period in the vicinity of $30\,$AU. It is, however, straightforward to apply this geometric correction, and weight the distribution accordingly \citep{2001AJ....121.2804B, 2004MNRAS.355..935M}.

In practice, we do not have a large aggregate of objects that were all discovered at the same distance, but rather a modest collection of objects, all discovered at different distances. Nevertheless, each of these detections amounts to an unbiased probe of the perihelion distribution interior to \textit{their} discovery distance, as described above. We can thus compare each discovery made at a particular distance to our modeled perihelion distributions for all objects interior to this discovery distance, in presence and in absence of P9.

\begin{figure}
\centering
\includegraphics[width=\columnwidth]{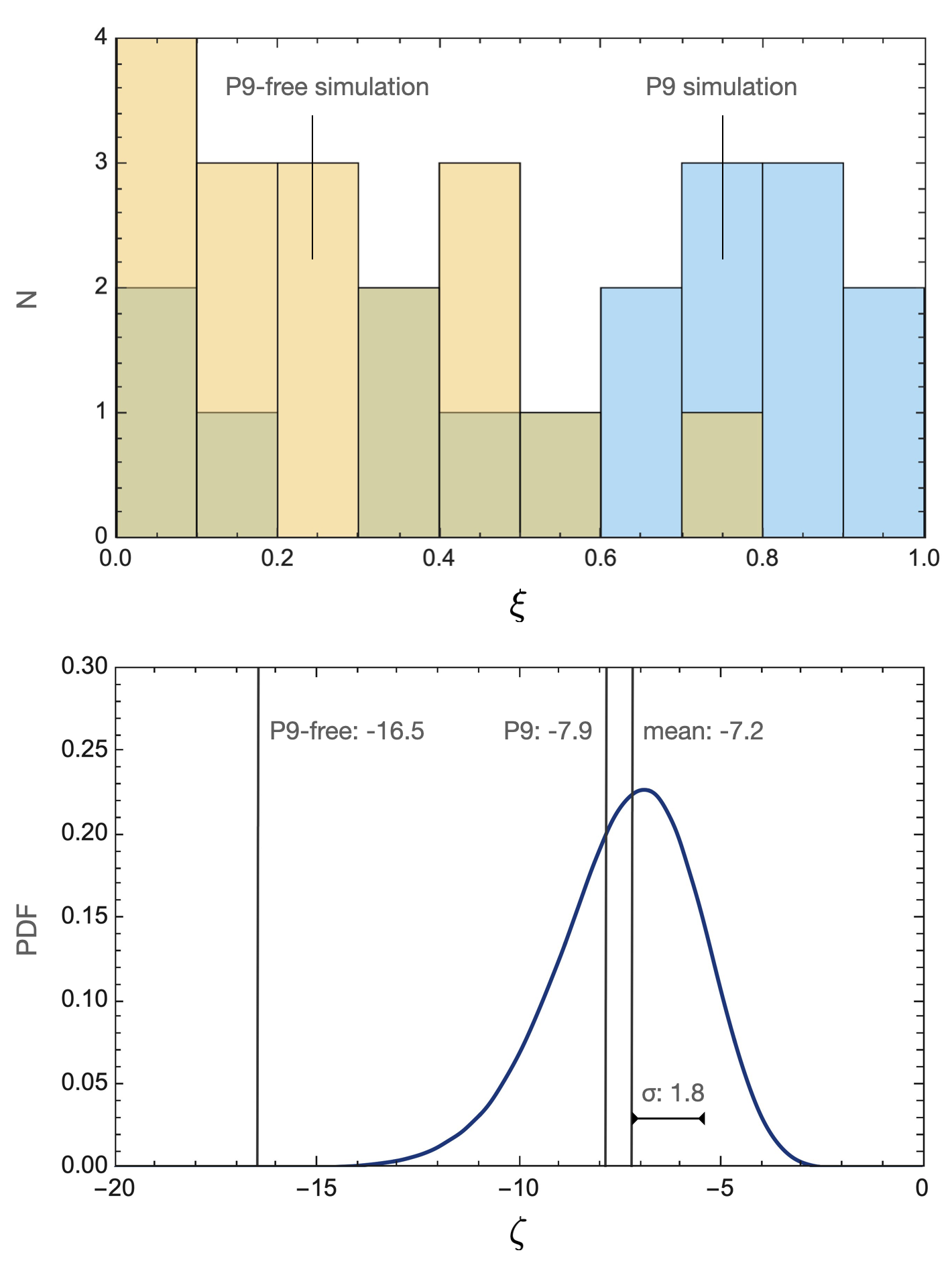}
\caption{Statistical comparison between the observed perihelion distribution of trans-Neptunian objects (TNOs) and simulations with and without Planet 9. The top panel illustrates the distribution of the variable $\xi$, which represents the cumulative distribution function values of the perihelia for the known objects, derived from the numerical models, accounting for observational bias. A uniform distribution of $\xi$ between 0 and 1 indicates a strong agreement between the observational data and the simulation. The P9 model shows a more uniform distribution of $\xi$ ($p = 0.41$) as opposed to the P9-free model ($p=0.0034$), suggesting a better match with the observational data. The bottom panel displays the probability distribution function of the logarithmic statistic, $\zeta$ which quantifies the uniformity of the $\xi$ distribution across the observed objects, adjusted for observational bias. The vertical lines represent the values of $\zeta$ corresponding to the P9 model (at $-7.9$) and the P9-free model (at $-16.5$), with the mean of the expected $\zeta-$distribution marked at $-7.2$. The spread of the distribution, denoted by $\sigma$ (standard deviation), is 1.8. The proximity of the P9 model statistic to the overall mean compared to the P9-free model -- lying more than $5\,\sigma$ away -- reflects a statistically significant preference for the Planet 9 hypothesis.}
\label{fig:newbias}
\end{figure}

Taking into account the geometric bias of the orbit, we use the simulation data to construct a PDF of the perihelion distribution for each \textit{discovery distance}, and then compute where each observed object falls within its individual CDF. We label the resulting quantity $\xi_j = \mathrm{CDF}_{r_j}(q_j)$. If the model perfectly matches the observations, $\xi_j$, should be uniformly distributed between 0 and 1. Any deviation from this uniformity serves as an unbiased statistical measure of the congruence between the model and the observational data. The top panel of Figure \ref{fig:newbias} shows the distribution of $\xi$ for the P9 and P9-free models. The Kolmogorov-Smirnov test for uniformity yields starkly distinct values, yielding $p = 0.41$ (P9) and $p=0.0034$ (P9-free), thus significantly favoring the model that includes P9. 

To further quantify the statistical discrepancy, we define the statistic $\zeta = \Pi_j^{N_{\rm{obj}}} \log(\xi_j)$. If the distribution of $\xi$ is uniform, then in the limit of $N_{\rm{obj}}\rightarrow \inf$, the distribution of expected values of $\zeta$ approaches a Gaussian. Though our sample size ($N_{\rm{obj}}=17$) is not sufficiently large for this result to hold exactly, this approach still allows for a rigorous measure of uniformity of $\xi$. We begin by constructing an expected distribution of $\zeta$ by computing $\Pi_j^{17} \log(\mathrm{U}(0,1))$ one million times, thereby generating a smooth histogram. The resulting curve is shown on the bottom panel of Figure \ref{fig:newbias}. We then compute the values of $\zeta$ for both simulations. Intriguingly, the P9 simulation's statistic ($\zeta=-7.9$) aligns closely with the mean of this distribution ($\langle\zeta\rangle=-7.2$), while the P9-free model's value ($\zeta=-16.5$) deviates significantly, falling approximately 5 standard deviations ($\sigma=1.8$) away from the peak. This comparison provides a quantitative assessment of the models, indicating a much higher likelihood of the P9 model given the current observational data.

As a check on the self-consistency of our statistical method, we conducted a validation exercise using simulated observations derived from a precisely defined synthetic distribution. This involved generating 100,000 orbits from on a synthetic distribution of TNOs characterized by a boxcar perihelion distribution within the 15 to 30 AU interval, a Gaussian distribution of inclinations with a dispersion of 15 degrees, and a semi-major axis distribution proportional\footnote{This distribution aligns with the steady-state solution emerging from a Fokker–Planck treatment of Neptune-scattering, where gravitational kicks are viewed as a diffusive process in specific energy.} to $dN/da\propto a^{-3/2}$. Subsequently, we simulated the observation of 20 objects using the OSSOS survey simulator \citep{2018FrASS...5...14L} and subjected these observations to our statistical analysis.

The outcomes of this test confirmed the method's robustness: applying our  metric to the synthetic population resulted in a KS $p$-value of 0.72 for the uniformity of $\xi$ and the value of $\zeta$ (-9.8) was only $0.6\,\sigma$ away from the mean value (-8.7) of the expected distribution. For completeness, we also acknowledge the potential for ecliptic longitude-dependent bias in the perihelion distribution due to the survey footprint of individual surveys like Pan-STARRS. Nevertheless, our simulations do not show any substantial dependence of the perihelion distribution on the longitude, meaning that this form of survey bias is almost certainly secondary, and is unlikely to meaningfully impact our conclusions.

\paragraph{Magnitude-limited bias correction} A distinct method to address observational bias in simulation data is the magnitude-limited correction approach. This technique simulates the detectability of objects within a generated orbital distribution for a survey with a specific limiting magnitude (see e.g., \citealt{2004MNRAS.355..935M}). While our current dataset comprises objects discovered across various surveys\footnote{The discovery magnitudes of the observational sample shown in Figure \ref{fig:data} range from approximately 21.5 to 24.5.} -- which precludes a straightforward application of this method -- understanding its implications remains beneficial. Specifically, this approach serves as a predictive tool for future uniform surveys, such as those planned for the Vera Rubin Observatory (VRO).

In choosing a limiting magnitude, we set $V_{\text{lim}} = 24$, coinciding with the anticipated capabilities of VRO. The size distribution assumes a power-law exponent of $\eta=2/3$ drawing on the results of \citet{2014ApJ...782..100F} for objects with absolute magnitudes range of $H\sim6-9$, though we find that the results are only weakly dependent on this choice. By accounting for the fraction of the orbit visible and the time spent traversing it as above, we generate biased distributions of $a$ and $q$ for both models and show them as histograms on Figure \ref{fig:orbdist}.

The biased semi-major axis distributions for both P9 and P9-free scenarios show a similar decay with increasing $a$, rendering them more akin to each other than in their unprocessed form. However, the perihelion distributions reveal a persistent, notable disparity: even after accounting for observational biases, the perihelion distribution in the P9-inclusive model retains a relatively flat distribution beyond approximately $16\,$AU. As already discussed above, this characteristic bears considerable resemblance to the distribution of the actual data presented in Figure \ref{fig:data}. In stark contrast, the P9-free model continues to exhibit a pronounced peak around $30\,$AU. This analysis indicates that the perihelion distribution of low-inclination TNOs is very likely to remain an important indicator for the existence of Planet 9 in forthcoming datasets.

\section{Discussion}

In this work, we have considered the orbital distribution of Neptune-crossing, low-inclination, long-period TNOs as a previously unidentified diagnostic for the existence of Planet 9. By conducting an $N-$body simulation of the solar system's long-term evolution, we have shown that P9-facilitated dynamics naturally drive orbits with $a>100\,$AU to Neptune-crossing eccentricities. Furthermore, we have devised a novel biasing procedure to compare simulation data with existing observations and demonstrated that the census of $q<30\,$AU TNOs strongly favors a model of the solar system that includes Planet 9.

\subsection{Observational Predictions}
As importantly as the comparison with existing observations, the results presented herein offer a set of readily-falsifiable predictions, with near-term prospects for resolution. To this end, we note that any comparison with the current data, even when biases are accounted for, is inherently imperfect, and a more uniformly acquired set of observations would provide a superior basis for testing our model. Fortunately, with the expected commencement of operations by the Vera Rubin Observatory, the orbital distribution of the class of objects considered here (Figure \ref{fig:data}) will come into much sharper focus, and the $a-q$ orbital distribution depicted in Figure \ref{fig:orbdist} will be tested directly.

\begin{figure}
\centering
\includegraphics[width=\columnwidth]{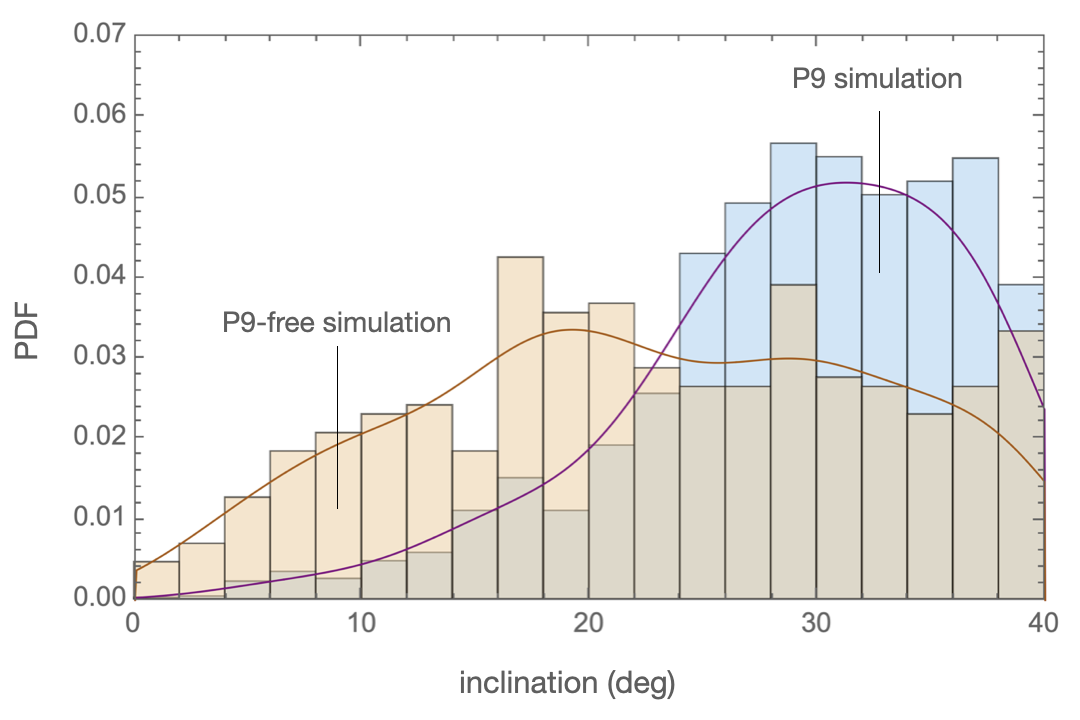}
\caption{Inclination distribution generated in presence and absence of Planet 9. The smooth curves represent the probability density function  fitted to the histogram data for each simulation. The P9 simulation shows a more pronounced peak and a distribution that extends towards higher inclinations, whereas the P9-free simulation exhibits a broader distribution with a gentler slope, peaking at lower inclinations. }
\label{fig:incdist}
\end{figure}

Another observational handle is provided by examining the absolute ratio of Neptune-crossing objects to those with $q>30\,$AU. Though the number of particles shown in the two panels of Figure \ref{fig:orbdist} is approximately equal, the number of \textit{total} orbital footprints that were used to generate the P9-free panel significantly exceeds that of the P9-inclusive panel. This discrepancy is a result of the more efficient injection of objects into the $q<30\,$AU region in the presence of Planet 9. More quantitatively, the ratio of Neptune-crossing objects with inclination $i < 40$ degrees and semi-major axis between $100$ and $1000\,$AU to those with $q>30\,$AU is $\sim3$\% in the Planet 9 scenario, compared to only $\sim0.5$\% in the P9-free case. While the current observational census does not provide a rigorous means to quantify this value, the advent of a comprehensive survey conducted by VRO will offer a more definitive opportunity to evaluate this prediction.


Finally, the predicted inclination distribution provides an avenue of inquiry. Although we have not delved into inclination biases in detail here it is noteworthy that the inclination distributions produced by P9 and P9-free simulations are strikingly different. Specifically, the distribution in the presence of P9 shows a steep rise with $i$, below 30 degrees. Meanwhile, the P9-free model exhibits a considerably flatter dispersion (Figure \ref{fig:incdist}). This prediction will be put on solid statistical footing with forthcoming results from VRO.

\subsection{Alternatives to Planet 9}
As concluding points, we note that while P9 explains the anomalous structure of the outer solar system in a unified framework, several alternative theories have been proposed to account for individual aspects of the P9 hypothesis. We briefly review these theories here and discuss how our work fits into this broader context. First, as already mentioned above, \citet{Nesv23} have shown how cluster-induced dynamics can generate the perihelion-detached population of TNOs, indicating that the $q-$broadening process may have been primordial, and that P9 is not strictly necessary to explain this observation\footnote{We note, however, that while the work of \citet{Nesv23} does not violate the constraints on the cluster properties imposed by the dynamical structure of the cold classical belt \citep{B20}, the parameters necessary to match the data are close to the upper limit of the allowed range.}. In a parallel vein, \citet{2022ApJ...938L..23H} have proposed the possibility that a few-Earth-mass rouge trans-Neptunian planet could have influenced the outer solar system's structure for hundreds of Myr, before being removed by some process\footnote{This removal process remains elusive because it is envisioned to occur after cluster dissipation and the work of \citet{2016ApJ...823L...3L} have shown that P9-type orbits have a negligible probability of being stripped away by passing stars within the field.}.

With respect to orbital clustering, \citet{2017AJ....154...50S, 2020PSJ.....1...28B}, and \citet{2021PSJ.....2...59N} have shown that individual surveys, which have examined limited areas of the sky, generally struggle to overcome their inherent observational biases sufficiently to rigorously determine the presence or absence of orbital alignment. This limitation has led some authors to interpret the observed orbital alignment as being illusory. Despite these challenges, it is  important to recognize that, even with the strong biases of the DES survey, \citet{2020PSJ.....1...28B} reported a $\sim2\,\sigma$ level of significance in the clustering of the longitude of ascending nodes. Additionally, a comprehensive observability analysis of all available data indicates that, after accounting for observational biases, distant KBOs are jointly clustered in Runge-Lenz (eccentricity) and angular momentum vectors with a significance level of approximately 99.6\% \citep{2017AJ....154...65B, 2019AJ....157...62B, 2021AJ....162..219B}. Finally, the observed anti-correlation between the rate of orbital diffusion and clustering within the data, (as discussed in \citealt{B19}), is unlikely to be attributable to observational bias alone. In a separate development, \citet{2023arXiv231020614H} have recently proposed that the alignment of three specific TNOs—Sedna, 2012 VP$_{113}$, and Leleakuhonua— is real, but is not created by P9. Instead, in their picture, these objects’ orbits could have been shaped by an early event in the solar system's history and have since precessed just enough to realign in the present epoch.

For the high-inclination population, \citet{Kaib19} have showed that the flux of retrograde Centaurs, as inferred from the OSSOS survey, is too large to be accounted for by a solar system model that excludes Planet 9. Their calculations further showed that the presence of Planet 9 could reconcile this discrepancy, although the adopted P9 parameters led to a median inclination of simulated detections that is a few degrees higher than the observed value. Still, as an alternative explanation, \citet{Kaib19} have also proposed that past migration of the Sun through the Galaxy could have enriched the Oort cloud, thereby enhancing the flux of retrograde Centaurs.

\medskip

In contrast to all of the above, the Neptune-crossing objects we have focused on in this work are distinct in a crucial way: due to their low inclinations and perihelia, these objects experience rapid orbital chaos and have short dynamical lifetimes (a simple \texttt{rebound} simulation illustrates that, in the absence of Planet 9, objects shown in Figure \ref{fig:data} have a dynamical lifetime on the order of $\sim100\,$Myr; \citealt{2015MNRAS.452..376R}). This implies that the dynamical process responsible for their current orbits is ongoing, not a relic of the distant past. Accordingly, any alternative to P9 that aims to explain this population must invoke active perturbations beyond those accounted for in calculations of \citet{Nesv23}. One such possibility is Modified Newtonian Dynamics (MOND), and recent proposals by \citet{2023AJ....166..168B} and \citet{2023MNRAS.525..805M} have suggested that MOND might explain some phenomena attributed to Planet 9. However, this hypothesis faces significant challenges, as \citet{2024arXiv240309555V} have shown that MOND significantly disrupts the observed specific energy distribution of long-period comets, and have further illustrated that certain variants of MOND fail to accurately account for the dynamics of detached objects. Additionally, \citet{2024MNRAS.527.4573B} have used wide binary data from the Gaia DR3 dataset to demonstrate that Newtonian gravity is very strongly favored over MOND on outer solar system scales. Equivalent conclusions were reached by \citet{2016arXiv160100947F,2023arXiv230301821F} from the vantage point of planetary ephemerides.

Another class of alternative theories involves the self-gravitational dynamics. \citet{2016MNRAS.457L..89M} were the first to propose the Inclination Instability as a mechanism for shaping the present-day architecture of the outer solar system. Follow-up studies have refined this picture, arguing that this instability could naturally manifest within a disk of initially planar but highly eccentric minor bodies, totaling around 10 Earth masses (see \citealt{2023ApJ...948L...1Z} and the references therein). Yet, as shown in the GPU-accelerated simulations of \citet{2023MNRAS.523.6103D}, even with such a massive disk, the inclination instability is fully suppressed if Neptune-scattering is modeled self-consistently. In an unrelated effort, \citet{2019AJ....157...59S} explored the idea of a similarly massive, mildly lopsided disk of planetesimals extending to about 700 AU as a potential driver of P9-like dynamics. Nevertheless, a suitable explanation for the origin of such a shepherding disk — and more importantly — its capacity to remain coherent on multi-Gyr timescales remains elusive.

In a study closely related to our work, \citet{Nesv23} showed that the same interplay between giant planet scattering and cluster-driven evolution that would have raised the perihelia of distant TNOs, could also trap as much as $\sim3$ Earth masses of material within the inner Oort Cloud — potentially facilitating non-trivial orbital evolution. In a recent paper \citep{BN24}, we analyzed the emergent phase-averaged dynamics of this scenario and found that the physical picture is qualitatively identical to that of von Zeipel–Lidov–Kozai (vZLK) cycles. Our calculations also indicate that unless the mass of the inner Oort Cloud is taken to be unreasonably large (i.e., tens to hundreds of Earth masses), the characteristic timescale of these cycles would far exceed the age of the sun. Therefore, at present, Planet 9 remains the only plausible explanation for the observed distribution of long-period Neptune-crossers.

\medskip

In summary, this work has introduced a new line of evidence supporting the Planet 9 hypothesis. Excitingly, the dynamics described here, along with all other lines of evidence for Planet 9, will soon face a rigorous test with the operational commencement of the Vera Rubin Observatory. This upcoming phase of exploration promises to provide critical insights into the mysteries of our solar system's outer reaches.

\acknowledgments We are thankful to Fred Adams, Gabriele Pichierri, Max Goldberg, and Juliette Becker for insightful discussions. We thank the anonymous referee for a thorough and insightful report that led to an improved manuscript. K.B. is grateful to Caltech and the David and Lucile Packard Foundation for their generous support.



\begin{thebibliography}


\bibitem[Adams(2010)]{Adams2010} Adams, F.~C.\ 2010, \araa, 48, 47. doi:10.1146/annurev-astro-081309-130830

\bibitem[Arakawa \& Kokubo(2023)]{2023A&A...670A.105A} Arakawa, S. \& Kokubo, E.\ 2023, \aap, 670, A105. doi:10.1051/0004-6361/202244743


\bibitem[Banik et al.(2024)]{2024MNRAS.527.4573B} Banik, I., Pittordis, C., Sutherland, W., et al.\ 2024, \mnras, 527, 4573. doi:10.1093/mnras/stad3393

\bibitem[Batygin \& Brown(2016a)]{BB2016a} Batygin, K. \& Brown, M.~E.\ 2016, \aj, 151, 22. doi:10.3847/0004-6256/151/2/22

\bibitem[Batygin \& Brown(2016b)]{BB2016b} Batygin, K. \& Brown, M.~E.\ 2016, \apjl, 833, L3. doi:10.3847/2041-8205/833/1/L3

\bibitem[Batygin \& Morbidelli(2017)]{BM17} Batygin, K. \& Morbidelli, A.\ 2017, \aj, 154, 229. doi:10.3847/1538-3881/aa937c

\bibitem[Batygin et al.(2019)]{B19} Batygin, K., Adams, F.~C., Brown, M.~E., et al.\ 2019, \physrep, 805, 1. doi:10.1016/j.physrep.2019.01.009

\bibitem[Batygin et al.(2020)]{B20} Batygin, K., Adams, F.~C., Batygin, Y.~K., et al.\ 2020, \aj, 159, 101. doi:10.3847/1538-3881/ab665d

\bibitem[Batygin \& Brown(2021)]{BB21} Batygin, K. \& Brown, M.~E.\ 2021, \apjl, 910, L20. doi:10.3847/2041-8213/abee1f

\bibitem[Batygin et al.(2021)]{2021ApJ...920..148B} Batygin, K., Mardling, R.~A., \& Nesvorn{\'y}, D.\ 2021, \apj, 920, 148. doi:10.3847/1538-4357/ac19a4

\bibitem[Batygin \& Nesvorn{\'y}(2024)]{BN24} Batygin, K. \& Nesvorn{\'y}, D. \ 2024, Celestial Mechanics and Dynamical Astronomy, \textit{submitted.}

\bibitem[Becker et al.(2017)]{2017AJ....154...61B} Becker, J.~C., Adams, F.~C., Khain, T., et al.\ 2017, \aj, 154, 61. doi:10.3847/1538-3881/aa7aa2

\bibitem[Becker et al.(2018)]{2018AJ....156...81B} Becker, J.~C., Khain, T., Hamilton, S.~J., et al.\ 2018, \aj, 156, 81. doi:10.3847/1538-3881/aad042

\bibitem[Bernardinelli et al.(2020)]{2020PSJ.....1...28B} Bernardinelli, P.~H., Bernstein, G.~M., Sako, M., et al.\ 2020, Planetary Science Journal, 1, 28. doi:10.3847/PSJ/ab9d80

\bibitem[Beust(2016)]{Beust2016} Beust, H.\ 2016, \aap, 590, L2. doi:10.1051/0004-6361/201628638

\bibitem[Brown(2001)]{2001AJ....121.2804B} Brown, M.~E.\ 2001, \aj, 121, 2804. doi:10.1086/320391

\bibitem[Brown et al.(2004)]{2004ApJ...617..645B} Brown, M.~E., Trujillo, C., \& Rabinowitz, D.\ 2004, \apj, 617, 645. doi:10.1086/422095

\bibitem[Brown \& Batygin(2016)]{BroBat16} Brown, M.~E. \& Batygin, K.\ 2016, \apjl, 824, L23. doi:10.3847/2041-8205/824/2/L23

\bibitem[Brown \& Batygin(2019)]{2019AJ....157...62B} Brown, M.~E. \& Batygin, K.\ 2019, \aj, 157, 62. doi:10.3847/1538-3881/aaf051

\bibitem[Brown(2017)]{2017AJ....154...65B} Brown, M.~E.\ 2017, \aj, 154, 65. doi:10.3847/1538-3881/aa79f4

\bibitem[Brown \& Batygin(2021)]{2021AJ....162..219B} Brown, M.~E. \& Batygin, K.\ 2021, \aj, 162, 219. doi:10.3847/1538-3881/ac2056

\bibitem[Brown \& Mathur(2023)]{2023AJ....166..168B} Brown, K. \& Mathur, H.\ 2023, \aj, 166, 168. doi:10.3847/1538-3881/acef1e


\bibitem[Chambers(1999)]{1999MNRAS.304..793C} Chambers, J.~E.\ 1999, \mnras, 304, 793. doi:10.1046/j.1365-8711.1999.02379.x

\bibitem[Clement \& Kaib(2020)]{2020AJ....159..285C} Clement, M.~S. \& Kaib, N.~A.\ 2020, \aj, 159, 285. doi:10.3847/1538-3881/ab9227

\bibitem[Clement \& Sheppard(2021)]{2021AJ....162...27C} Clement, M.~S. \& Sheppard, S.~S.\ 2021, \aj, 162, 27. doi:10.3847/1538-3881/abfe07


\bibitem[Das \& Batygin(2023)]{2023MNRAS.523.6103D} Das, A. \& Batygin, K.\ 2023, \mnras, 523, 6103. doi:10.1093/mnras/stad1687



\bibitem[Fienga et al.(2016)]{2016arXiv160100947F} Fienga, A., Laskar, J., Manche, H., et al.\ 2016, arXiv:1601.00947. doi:10.48550/arXiv.1601.00947

\bibitem[Fienga \& Minazzoli(2023)]{2023arXiv230301821F} Fienga, A. \& Minazzoli, O.\ 2023, arXiv:2303.01821. doi:10.48550/arXiv.2303.01821

\bibitem[Fraser et al.(2014)]{2014ApJ...782..100F} Fraser, W.~C., Brown, M.~E., Morbidelli, A., et al.\ 2014, \apj, 782, 100. doi:10.1088/0004-637X/782/2/100

\bibitem[Fouchard et al.(2014)]{2014Icar..231..110F} Fouchard, M., Rickman, H., Froeschl{\'e}, C., et al.\ 2014, \icarus, 231, 110. doi:10.1016/j.icarus.2013.11.032


\bibitem[Gladman et al.(2009)]{2009ApJ...697L..91G} Gladman, B., Kavelaars, J., Petit, J.-M., et al.\ 2009, \apjl, 697, L91. doi:10.1088/0004-637X/697/2/L91

\bibitem[Gomes et al.(2005)]{2005Natur.435..466G} Gomes, R., Levison, H.~F., Tsiganis, K., et al.\ 2005, \nat, 435, 466. doi:10.1038/nature03676

\bibitem[Gomes et al.(2015)]{2015Icar..258...37G} Gomes, R.~S., Soares, J.~S., \& Brasser, R.\ 2015, \icarus, 258, 37. doi:10.1016/j.icarus.2015.06.020



\bibitem[Hadden et al.(2018)]{2018AJ....155..249H} Hadden, S., Li, G., Payne, M.~J., et al.\ 2018, \aj, 155, 249. doi:10.3847/1538-3881/aab88c

\bibitem[Heisler \& Tremaine(1986)]{1986Icar...65...13H} Heisler, J. \& Tremaine, S.\ 1986, \icarus, 65, 13. doi:10.1016/0019-1035(86)90060-6

\bibitem[Huang et al.(2022)]{2022ApJ...938L..23H} Huang, Y., Gladman, B., Beaudoin, M., et al.\ 2022, \apjl, 938, L23. doi:10.3847/2041-8213/ac9480

\bibitem[Huang \& Gladman(2023)]{2023arXiv231020614H} Huang, Y. \& Gladman, B.\ 2023, arXiv:2310.20614. doi:10.48550/arXiv.2310.20614


\bibitem[Izidoro et al.(2023)]{2023DPS....5550009I} Izidoro, A., Raymond, S., Kaib, N., et al.\ 2023, \dps



\bibitem[Kaib et al.(2011)]{2011Icar..215..491K} Kaib, N.~A., Ro{\v{s}}kar, R., \& Quinn, T.\ 2011, \icarus, 215, 491. doi:10.1016/j.icarus.2011.07.037

\bibitem[Kaib et al.(2019)]{Kaib19} Kaib, N.~A., Pike, R., Lawler, S., et al.\ 2019, \aj, 158, 43. doi:10.3847/1538-3881/ab2383

\bibitem[Khain et al.(2018)]{Kain18} Khain, T., Batygin, K., \& Brown, M.~E.\ 2018, \aj, 155, 250. doi:10.3847/1538-3881/aac212

\bibitem[Khain et al.(2020)]{2020PASP..132l4401K} Khain, T., Becker, J.~C., \& Adams, F.~C.\ 2020, \pasp, 132, 124401. doi:10.1088/1538-3873/abbd8a


\bibitem[Lawler et al.(2018)]{2018FrASS...5...14L} Lawler, S.~M., Kavelaars, J.~J., Alexandersen, M., et al.\ 2018, Frontiers in Astronomy and Space Sciences, 5, 14. doi:10.3389/fspas.2018.00014


\bibitem[Li \& Adams(2016)]{2016ApJ...823L...3L} Li, G. \& Adams, F.~C.\ 2016, \apjl, 823, L3. doi:10.3847/2041-8205/823/1/L3

\bibitem[Li et al.(2018)]{2018AJ....156..263L} Li, G., Hadden, S., Payne, M., et al.\ 2018, \aj, 156, 263. doi:10.3847/1538-3881/aae83b


\bibitem[Madigan \& McCourt(2016)]{2016MNRAS.457L..89M} Madigan, A.-M. \& McCourt, M.\ 2016, \mnras, 457, L89. doi:10.1093/mnrasl/slv203

\bibitem[Migaszewski(2023)]{2023MNRAS.525..805M} Migaszewski, C.\ 2023, \mnras, 525, 805. doi:10.1093/mnras/stad2250

\bibitem[Millholland \& Laughlin(2017)]{2017AJ....153...91M} Millholland, S. \& Laughlin, G.\ 2017, \aj, 153, 91. doi:10.3847/1538-3881/153/3/91

\bibitem[Morbidelli et al.(2004)]{2004MNRAS.355..935M} Morbidelli, A., Emel'yanenko, V.~V., \& Levison, H.~F.\ 2004, \mnras, 355, 935. doi:10.1111/j.1365-2966.2004.08372.x

\bibitem[Morbidelli et al.(2005)]{2005Natur.435..462M} Morbidelli, A., Levison, H.~F., Tsiganis, K., et al.\ 2005, \nat, 435, 462. doi:10.1038/nature03540


\bibitem[Napier et al.(2021)]{2021PSJ.....2...59N} Napier, K.~J., Gerdes, D.~W., Lin, H.~W., et al.\ 2021, Planetary Science Journal, 2, 59. doi:10.3847/PSJ/abe53e

\bibitem[Nesvorn{\'y} et al.(2017)]{2017ApJ...845...27N} Nesvorn{\'y}, D., Vokrouhlick{\'y}, D., Dones, L., et al.\ 2017, \apj, 845, 27. doi:10.3847/1538-4357/aa7cf6

\bibitem[Nesvorn{\'y}(2018)]{NesvReview} Nesvorn{\'y}, D.\ 2018, \araa, 56, 137. doi:10.1146/annurev-astro-081817-052028

\bibitem[Nesvorn{\'y} et al.(2023)]{Nesv23} Nesvorn{\'y}, D., Bernardinelli, P., Vokrouhlick{\'y}, D., et al.\ 2023, \icarus, 406, 115738. doi:10.1016/j.icarus.2023.115738



\bibitem[Oldroyd \& Trujillo(2021)]{2021AJ....162...39O} Oldroyd, W.~J. \& Trujillo, C.~A.\ 2021, \aj, 162, 39. doi:10.3847/1538-3881/abfb6f


\bibitem[Press et al.(1992)]{1992nrfa.book.....P} Press, W.~H., Teukolsky, S.~A., Vetterling, W.~T., et al.\ 1992, Cambridge: University Press, |c1992, 2nd ed.



\bibitem[Rein \& Tamayo(2015)]{2015MNRAS.452..376R} Rein, H. \& Tamayo, D.\ 2015, \mnras, 452, 376. doi:10.1093/mnras/stv1257


\bibitem[Saillenfest et al.(2017)]{2017CeMDA.129..329S} Saillenfest, M., Fouchard, M., Tommei, G., et al.\ 2017, Celestial Mechanics and Dynamical Astronomy, 129, 329. doi:10.1007/s10569-017-9775-7

\bibitem[Sefilian \& Touma(2019)]{2019AJ....157...59S} Sefilian, A.~A. \& Touma, J.~R.\ 2019, \aj, 157, 59. doi:10.3847/1538-3881/aaf0fc

\bibitem[Shankman et al.(2017)]{2017AJ....154...50S} Shankman, C., Kavelaars, J.~J., Bannister, M.~T., et al.\ 2017, \aj, 154, 50. doi:10.3847/1538-3881/aa7aed



\bibitem[Thomas \& Morbidelli(1996)]{1996CeMDA..64..209T} Thomas, F. \& Morbidelli, A.\ 1996, Celestial Mechanics and Dynamical Astronomy, 64, 209. doi:10.1007/BF00728348

\bibitem[Trujillo \& Sheppard(2014)]{2014Natur.507..471T} Trujillo, C.~A. \& Sheppard, S.~S.\ 2014, \nat, 507, 471. doi:10.1038/nature13156

\bibitem[Tsiganis et al.(2005)]{2005Natur.435..459T} Tsiganis, K., Gomes, R., Morbidelli, A., et al.\ 2005, \nat, 435, 459. doi:10.1038/nature03539




\bibitem[Vokrouhlick{\'y} et al.(2024)]{2024arXiv240309555V} Vokrouhlick{\'y}, D., Nesvorn{\'y}, D., \& Tremaine, S.\ 2024, arXiv:2403.09555. doi:10.48550/arXiv.2403.09555



\bibitem[Wiegert \& Tremaine(1999)]{1999Icar..137...84W} Wiegert, P. \& Tremaine, S.\ 1999, \icarus, 137, 84. doi:10.1006/icar.1998.6040





\bibitem[Zderic \& Madigan(2023)]{2023ApJ...948L...1Z} Zderic, A. \& Madigan, A.-M.\ 2023, \apjl, 948, L1. doi:10.3847/2041-8213/accde2



\end{thebibliography}

\bibliographystyle{aasjournal}

\end{document}